\documentclass[10pt,conference]{IEEEtran}
\IEEEoverridecommandlockouts

\usepackage{cite}
\usepackage{amsmath,amssymb,amsfonts}
\usepackage{graphicx}
\usepackage{textcomp}
\usepackage{xcolor}
\usepackage{booktabs}
\usepackage{url}
\usepackage{hyperref}
\usepackage{tcolorbox}
\usepackage{listings}

\def\BibTeX{{\rm B\kern-.05em{\sc i\kern-.025em b}\kern-.08em
    T\kern-.1667em\lower.7ex\hbox{E}\kern-.125emX}}

\usepackage{eso-pic}
\AddToShipoutPictureFG*{\AtPageLowerLeft{\put(\LenToUnit{0.08\paperwidth},\LenToUnit{0.022\paperheight}){\parbox{0.84\paperwidth}{\scriptsize\centering \textcopyright~2026 IEEE. Personal use of this material is permitted. Permission from IEEE must be obtained for all other uses, in any current or future media, including reprinting/republishing this material for advertising or promotional purposes, creating new collective works, for resale or redistribution to servers or lists, or reuse of any copyrighted component of this work in other works.}}}}

\begin{document}

\title{Which Link Should I Look At First?: Automated Link Ranking in Code Reviews}

\author{\IEEEauthorblockN{Panya Trakoolgerntong\IEEEauthorrefmark{1}, Arus Thienmee\IEEEauthorrefmark{1}, Patsathorn Chadbantern\IEEEauthorrefmark{1}, Chaiyong Ragkhitwetsagul\IEEEauthorrefmark{2},\\ Morakot Choetkiertikul\IEEEauthorrefmark{1}, Tao Xiao\IEEEauthorrefmark{3}, Patanamon Thongtanunam\IEEEauthorrefmark{4}, Yasutaka Kamei\IEEEauthorrefmark{3}}
\IEEEauthorblockA{\IEEEauthorrefmark{1}Faculty of Information and Communication Technology, Mahidol University, Nakhon Pathom, Thailand\\
\IEEEauthorrefmark{2}Department of Computer Engineering, Faculty of Engineering, Chulalongkorn University, Bangkok, Thailand\\
\IEEEauthorrefmark{3}Graduate School and Faculty of Information Science and Electrical Engineering, Kyushu University, Fukuoka, Japan}
\IEEEauthorrefmark{4}School of Computing and Information Systems, the University of Melbourne, Melbourne, Australia}

\maketitle

\begin{abstract}
Pull requests (PRs) on GitHub often include hyperlinks to issues, commits, and documentation that reviewers must manually inspect, which can lead to context switching and increased cognitive load. We conduct an empirical study of 486,423 PRs from 50 popular GitHub repositories and find that hyperlinks are pervasive, with an average of 4.69 links per PR description and a highly skewed distribution. Motivated by these findings, we propose an automated approach that ranks the links within a PR description by their contextual relevance. We compare techniques spanning lexical, semantic, and supervised methods, including TF-IDF, large language model (LLM)-based encoders, cross-encoder rerankers, and Learning-to-Rank (LTR). On a manually annotated dataset of 700 high-link-density PRs, our LTR model achieves the best ranking performance. We implement the approach as a GitHub Action, MonnetalX, which presents the ranked links together with summaries of the top-ranked resources. A human evaluation involving 14 participants shows that the ranked links help reviewers save time, stay focused, and reduce cognitive load, with 85.7\% of participants rating the reduction in cognitive load as good or excellent. Our results show that links are pervasive in PR descriptions. LTR with lexical and semantic features performs best for link ranking, and link ranking benefits software developers by saving time, enabling greater focus, and reducing cognitive load. 
\end{abstract}


\section{Introduction}
\label{sec:intro}

In modern collaborative software development, pull requests (PRs) play a central role in facilitating code review, discussion, and quality assurance~\cite{Gousios2014}. PR descriptions, a dedicated section where a developer explains the context, purpose, and impact of their proposed code changes before those changes are merged into the main codebase, increasingly contain numerous external links, such as documentation, design references, issue trackers, and related commits, that reviewers must navigate to fully understand the proposed changes. This reliance on dispersed, multi-source information creates a substantial cognitive burden, leads to information overload, and slows the review process~\cite{Bacchelli,Wang2021}. 

Prior work confirms that link sharing is a common practice in code review and serves diverse purposes such as providing context and elaborating on technical details~\cite{Wang2021}. 
Including links to related issues is also considered one of the best practices for writing high-quality PR descriptions~\cite{Jiang2026}. However, not all links are equally relevant: a PR may reference dozens of resources, only a few of which are essential for understanding the change. When reviewers must inspect every link to identify the important ones, the very mechanism intended to provide context instead becomes a source of overhead~\cite{trakoolgerntong}. 

\begin{figure}[tb]
    \centering
    \includegraphics[width=\columnwidth]{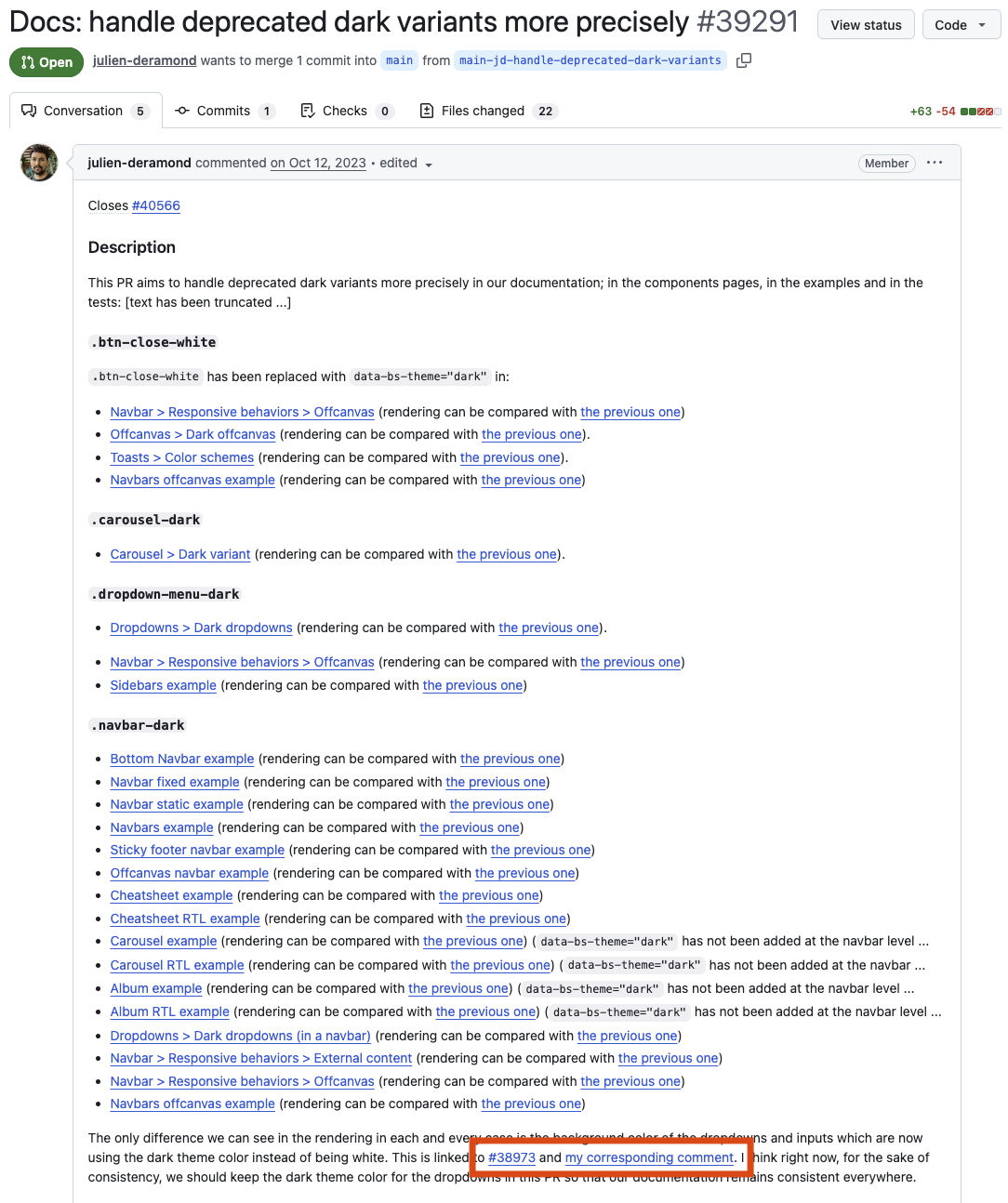}
    \caption{A PR with multiple links. Two highly relevant links are at the bottom (highlighted in red). The figure was edited to fit within the limited space.}
    \label{fig:example}
\end{figure}

As a motivating example, Figure~\ref{fig:example} shows a PR from the Bootstrap project, a framework for developing responsive, mobile-first web projects,\footnote{https://github.com/twbs/bootstrap/pull/39291} that addresses handling the deprecated dark variant UI and contains 59 links. While most links point to deployed previews and documentation, 2 links point to the issue that led to this PR, and a comment (from the PR creator) in that issue explains the problem and the solution. They are important for understanding the context of this PR and are placed almost at the end (the 50th and 51st links). Thus, the reviewers of the PR must go through several linked resources before reaching these two important links.

While existing studies empirically investigate the importance of shared links in PR, few have addressed the challenge of prioritizing them.
First, empirical studies of code review have characterized how and why links are shared. Wang et al.~\cite{Wang2021} found that, in Gerrit code review platforms, roughly one quarter of code reviews contain at least one shared link.
Jiang et al.~\cite{jiang2017linksharing} similarly observed on GitHub that links in review comments are nearly evenly split between internal and external resources, and that PRs whose review comments contain links attract more comments and take longer to review. %
Second, a growing body of work reduces reviewer effort through automated summarization: fine-tuned generative models produce PR titles~\cite{irsan2022autoprtitletoolautomaticpull, zhang2022} and PR descriptions~\cite{Kuang2021, Sakib2024}, data cleaning further improves the quality of generated descriptions~\cite{tire2025evaluatingimpactdatacleaning}, and, most recently, Trakoolgerntong et al.~\cite{trakoolgerntong} summarized the linked resources using an LLM, showing that LLM-generated link summaries improve substantially when structured context is provided. Third, links between software artifacts have been studied as a traceability problem: several techniques recover or predict missing issue-PR links~\cite{alshara2023mlaugmented, bai2024aipl, jiang2025linkfinder} or trace release notes back to PRs and issues~\cite{nath2025traceability}, and learning-to-rank has been applied at the PR level to prioritize which PRs reviewers should examine first~\cite{zhao2019ltr}.
However, none of these approaches addresses the relevance of the links \textit{within} a PR description. 
As a result, reviewers are still left to manually go through dozens of links one by one.  

This gap matters because link prioritization in PR descriptions can reduce the time required for downstream use of linked content, including a summary, a preview, or a recommendation, which is only as useful as the relevance of the links it is built upon. To the best of our knowledge, no prior work has addressed the problem of ranking links in PR descriptions. This paper closes this gap by \textit{automatically ranking the links embedded in a PR description by their contextual relevance, so that reviewers can focus on the resources that matter most}.

We examine the prevalence of links in PR descriptions and the effectiveness of link-ranking techniques in a three-phase study. First, we conduct a large-scale empirical analysis of 486,423 PRs across 50 popular GitHub repositories to characterize the prevalence, types, and distribution of links in PR descriptions. We find that links are pervasive, with an average of 4.69 links per PR description and a highly skewed distribution: 75\% of PRs contain at most 3 links, while the remaining 25\% contain more than 3 and up to 1,571 links.
Then, we conduct a comparative evaluation of link ranking techniques spanning lexical, semantic, and supervised approaches, including TF-IDF, LLM-based encoders, cross-encoder rerankers, and gradient-boosted learning-to-rank (LTR) models. Using a manually annotated dataset of 700 PRs with extensive links, we find that an LTR model that combines lexical, semantic, and structural features achieves the best ranking performance, ordering links significantly closer to the ground truth than purely semantic alternatives. 

To assess whether this link ranking offers value during code reviews, we implement the approach as a GitHub Action (called MonnetalX~\cite{monnetalx}) that surfaces the ranked links (together with summaries of the top-ranked resources), and we evaluate it with qualified participants. We found that the ranked links help reviewers save time, stay focused, and reduce cognitive load, with 85.7\% of participants rating the reduction in cognitive load as good or excellent.

This study makes the following contributions:
\begin{itemize}
    \item A large-scale preliminary study of 486,423 PRs across 50 popular GitHub repositories, characterizing hyperlink prevalence, types, and distribution, and establishing that link density is highly skewed.
    \item A comparative evaluation of link ranking techniques, including TF-IDF, LLM-based encoders, cross-encoder rerankers, and a LightGBM LambdaMART model trained on a manually annotated dataset of 700 PRs, identifying the feature combination that yields the most effective ranking.
    \item A GitHub Action tool, MonnetalX, that automatically ranks and summarizes links in PR descriptions.
\end{itemize}

\section{Background and Related Work}
\label{sec:background}

\subsection{Code Review and Link Sharing in PRs}

Code review, where independent developers manually inspect source code written by others, is a widely adopted practice in modern software development.
McIntosh et al.~\cite{McIntosh2014} report that high code review coverage and participation are significantly linked to fewer post‑release defects.
In addition, code review is not only about finding defects but also about improving quality, maintainability, and knowledge sharing~\cite{Bacchelli,Badampudi2023}.  GitHub has popularized the pull request model for code review, where developers propose and discuss changes~\cite{Gousios2014}. 

Wang et al.~\cite{Wang2021} conducted a mixed-method study of code reviews from OpenStack and Qt, finding that about one quarter of reviews contain at least one link, that internal links mainly provide context while external links elaborate on technical details, and that more internal links are weakly correlated with longer review times. Jiang et al.~\cite{jiang2017linksharing} analyzed PRs from ten GitHub projects and found that links in review comments are nearly evenly split between internal (51.49\%) and external (48.51\%) resources, and that PRs with links receive more comments and take longer to review, similar to Wang et al.'s study~\cite{Wang2021}. Li et al.~\cite{li2018linking} analyzed issues and PRs of Python projects and identified six kinds of linking relationships in GitHub issues and PRs, e.g., a dependent relationship where an issue/PR's resolution depends on another one or a duplicate relationship where two issues/PRs address the same problem. These studies establish that link sharing is valuable yet costly for reviewers, but none assess or rank the relevance of links.

\subsection{Artifact Link Recovery and Ranking in PR}

A separate line of research treats links between artifacts as a traceability problem and establishes links that developers omitted. Alshara et al.~\cite{alshara2023mlaugmented} recovered missing issue--PR links via clustering over lexical and semantic similarity, which shows high accuracy on the PI-Link dataset~\cite{Alshara2023} containing links between GitHub pull requests and issues; Bai et al.~\cite{bai2024aipl} predicted such links with knowledge-aware heterogeneous graph learning; and Jiang et al.~\cite{jiang2025linkfinder} recommended linked issues for a PR with LinkFinder. Nath et al.~\cite{nath2025traceability} used LLMs to trace release notes back to PRs, commits, and issues. 

Ranking has been applied to some of the review artifacts. Zhao et al.~\cite{zhao2019ltr} used learning-to-rank to order incoming PRs by their likelihood of being quickly merged or rejected, outperforming FIFO and small-size-first baselines. 
Fregnan et al.~\cite{Fregnan2022} and Bagirov et al. \cite{Bagirov2023} demonstrate that the alphabetical ordering of files in code review introduces a cognitive bias, showing that files presented earlier in a PR receive significantly more comments than those shown later. 

These studies show that the ordering of code review artifacts, such as PRs or files within a PR, matters. Nonetheless, none of them has looked into the ordering of links within a PR description.
Outside code review, ranking software artifacts by their similarity to a query is well established; for example, BIKER~\cite{huang2018biker} retrieves Stack Overflow questions similar to a programming task and ranks API methods by their similarity to the task and the API documentation. Unlike such retrieval approaches, which select candidates from a large corpus, we order the links that an author has already placed in a PR description.

\subsection{PR and Link Summarization}

Several studies reduce reviewer effort by summarizing PR content. 
Irsan et al.~\cite{irsan2022autoprtitletoolautomaticpull} fine-tuned BART for automatic PR title generation based on the pull request description, commit messages, and the associated issue titles. Zhang et al.~\cite{zhang2022} proposed a sequence-to-sequence model with attention and copy mechanisms. %
Closest to our work, Trakoolgerntong et al.~\cite{trakoolgerntong} introduced AILinkPreviewer, the first automated tool for summarizing the hyperlinks embedded in PR descriptions.\footnote{\url{https://github.com/c4rtune/AILinkPreviewer}} Evaluating several LLM prompt configurations, they showed that structured contextual inputs (link metadata, surrounding text, and extracted content) substantially improve summary quality, with structured chain-of-thought prompting performing best. 
Nonetheless, all of these approaches either summarize the PR's content and treat embedded links as plain text or discard them entirely.
Moreover, summarization alone does not determine which links deserve the reviewer's attention in code review. Our ranking approach offers this missing prioritization.

\section{Research Questions}
\label{sec:research_questions}
We aim to answer the following research questions:
\begin{itemize}
    \item \textbf{RQ1:} \textit{How prevalent are hyperlinks in PR descriptions, and what is their distribution across different types of resources?} Understanding the landscape of hyperlink usage in PR descriptions is critical for motivating and designing an effective ranking approach. This question is addressed through a large-scale analysis of PR descriptions across multiple repositories, characterizing link frequency, types (e.g., internal vs.\ external), and distribution patterns.
    \item \textbf{RQ2:} \textit{What is the effectiveness of different techniques in ranking links in PR descriptions based on their relevance?} We evaluate the performance of several link ranking techniques on a manually annotated dataset of PR descriptions, including TF-IDF, LLM-based encoders, cross-encoder rerankers, and learning-to-rank models. This comparative analysis will identify the most effective approach for ordering links by contextual relevance, which is essential for reducing reviewer cognitive load.
    \item \textbf{RQ3:} \textit{How do software developers perceive the effectiveness of the link ranking approach during PR reviews?} A human evaluation will assess the practical impact of our approach on the PR review process, focusing on perceived effectiveness in helping reviewers quickly understand linked resources.
\end{itemize}

\begin{figure*}[tb]
    \centering
    \includegraphics[width=\textwidth]{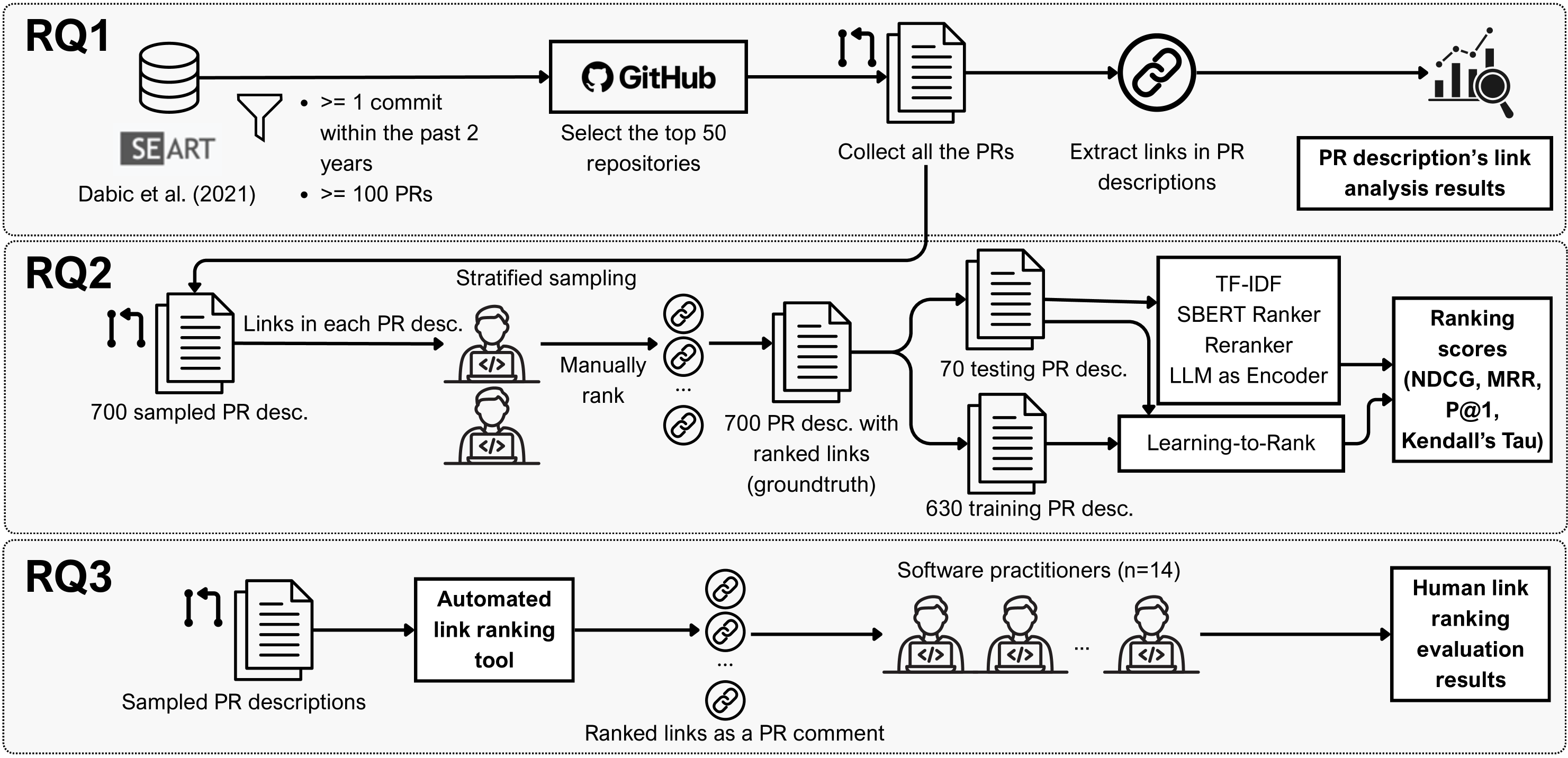}
    \caption{The overview of the study}
    \label{fig:overview}
\end{figure*}

\section{Methodology}
\label{sec:methodology}

The overview of our study is depicted in Figure~\ref{fig:overview}. It is divided into 3 phases, each addressing 1 of the 3 research questions. We explain each phase in detail below.

\subsection{RQ1: Hyperlink Usage in PRs}
\label{sec:rq1_methodology}

To understand the prevalence of links in GitHub pull request descriptions, we analyzed pull requests in popular GitHub repositories. The list of repositories is retrieved from the SEART GitHub Search platform~\cite{Dabic2021}, which provides advanced filtering and metadata for large-scale analysis of GitHub repositories. Repositories were first sorted by star rating to prioritize widely used, actively maintained projects. We applied two inclusion criteria used in the previous study~\cite{trakoolgerntong}: (1) the repository must have received a commit within the past two years (from 20 May 2024 backward), ensuring the project remains active, and (2) the repository must contain at least 100 pull requests, guaranteeing sufficient review activity for empirical analysis. From the resulting pool, we selected the top 50 repositories that qualify as engineered software projects~\cite{Munaiah2017}, defined as projects that adhere to established software engineering practices, such as systematic documentation, testing, and project management.

To systematically collect pull requests with links, we developed a custom extraction script using the GitHub GraphQL API. The script retrieved pull requests from each selected repository using a paginated query to avoid memory overload and to ensure complete coverage of historical PR data. For every pull request, we extracted the PR number, title, and body text, and then applied a regular expression-based parser \verb|https?://[^\s<>()\[\]`]+| on the body text to locate the links in the PR description.

To identify all visible HTTP and HTTPS links referenced within the PR description, each extracted link was subsequently analyzed to determine its media type (e.g., text, image, video, audio) by issuing an HTTP HEAD request and inspecting the returned Content-Type header. 
We analyzed 486,423 pull requests, totaling 2,281,756 links in their descriptions. 

\subsection{RQ2: Evaluation of Link Ranking Models}
\label{sec:rq2_methodology}

We explored several techniques for ranking links in GitHub PR descriptions. Across all techniques, the PR description text is used as-is, and each link's content is retrieved and used for the ranking. We discuss each of them as follows:

\subsubsection{Term Frequency-Inverse Document Frequency (TF-IDF)}

TF-IDF is a vector space model that scores the importance of a word to a document relative to a corpus~\cite{tf-idf}. It relies on lexical statistics and does not capture semantic meaning beyond exact term matching. For each GitHub repository, we build one TF-IDF model using all of that repository's PR descriptions as the corpus, so the representation reflects the local vocabulary and domain-specific terminology of each project. To rank the links, the PR description (the query) and each link's content (a candidate document) are vectorized with the repository's TF-IDF model, and the links are ranked by the cosine similarity between the description vector and each link vector, following prior work on traceability link recovery~\cite{le2015rclinker}, bug recommendation~\cite{yang2016combining}, and tag recommendation~\cite{xia2013tag}.

\subsubsection{Large Language Models as Encoders}

LLMs can serve as semantic encoders that transform text into dense vector representations, capturing relatedness even when the PR description and linked content share little vocabulary, a setting where lexical methods such as TF-IDF suffer from vocabulary mismatch~\cite{karpukhin2020dense}. We employ two LLM encoders, \textit{Mistral-7B} and \textit{DeepSeek-Coder}, which are widely used and maximize parameter capacity while remaining supported by our experimental infrastructure. To rank the links, we encode the PR description and each link's content and use their cosine similarity as the relevance estimate.

\subsubsection{Sentence-BERT Embedder} The Sentence-BERT (SBERT)~\cite{sentrans} uses a Bi-Encoder architecture that encodes queries and documents independently into dense vectors compared via cosine similarity, enabling fast retrieval at the cost of the fine-grained precision of cross-encoders. We selected the widely used \textit{all-MiniLM-L6-v2} model, which maps text into a 384-dimensional space at low computational cost while achieving strong performance on semantic-similarity benchmarks. To rank the links, we embed the PR description and each link and sort the links by their cosine similarity to the description.

\subsubsection{Reranker (Cross-Encoder)}
Rerankers are LLM-based ranking models with built-in encoders and scoring. Unlike embedding models that use a Bi-Encoder (such as SBERT), rerankers use a Cross-Encoder architecture, sacrificing recall for higher precision. We chose \textit{jina-reranker-v3}~\cite{Wang2025}, a 0.6B-parameter multilingual reranker with a last-but-not-late interaction that encodes documents separately and compares each candidate at the end of encoding, so the model's understanding expands to every document and query. To rank the links, we feed the PR description and each link's content to the model's built-in predict function, which returns a relevance score for each link, and then sort the links by these scores.

\subsubsection{Learning-to-rank}
\label{sec:ltr}
Learning-to-rank (LTR) models are a class of supervised machine learning algorithms designed to optimize the ordering of items (in this case, links) based on their relevance to a given query (the PR description). LTR models learn from labeled training data, where each query-document pair is annotated with a relevance score. The model then predicts the relevance of new query-document pairs and ranks them accordingly. 
In our context, we frame link ranking as an LTR problem. Given a PR description and a set of extracted links' content, the goal is to order links by their contextual relevance to the PR description. 
Using LTR, we trained a \textit{LambdaMART} model~\cite{lambdamart} with \textit{LightGBM gradient boosting} framework~\cite{lightGBM} using two feature configurations:

\begin{itemize}
    \item \textbf{4-feature set:} This set of features includes (1) cosine similarity between PR and link content (computed using all-MiniLM-L6-v2 SBERT embeddings~\cite{sentrans}), (2) a flag indicating whether the linked resource is from the same repository as the current PR (1 if from the same repository, 0 otherwise), (3) a flag indicating whether the linked resource is created by the assigned reviewer of the current PR (1 if the linked resource, e.g., PR, issue, commit is created by the assigned reviewer, 0 otherwise), and (4) a flag indicating whether the link resource is a code change, i.e., diff (1 for code changes, 0 otherwise).
    \item \textbf{8-feature set:} Extends the 4-feature set with link content length, i.e., the number of words in the linked resource's content, and unigram, bigram, and trigram lexical overlap between PR description and links' content.
\end{itemize}

The features capture complementary relevance signals. Cosine similarity captures the semantic relatedness of a link to the PR. The same-repository flag reflects that internal resources, such as issues, PRs, and code of the same project, usually provide context for the change. The reviewer flag reflects that a reviewer may already be familiar with, or find most relevant, the resources they created. The code-change flag marks diffs, which relate directly to the change under review. The $n$-gram overlaps add lexical evidence, such as shared identifiers and terms, that embeddings may blur.
Link content length reflects how much information a linked resource offers: longer resources, such as detailed issue reports or design documents, tend to carry more context for the change. It also indirectly signals inaccessible resources, since a link whose content cannot be retrieved is represented by a short, fixed placeholder text.
For the lexical features, both texts are lowercased and split on whitespace, without stemming or stop-word removal. The $n$-gram overlap is the fraction of the link content's $n$-grams that also occur in the PR description, $|G_n(\ell) \cap G_n(d)| \,/\, \max(|G_n(\ell)|, 1)$ for $n \in \{1, 2, 3\}$, where $G_n(\cdot)$ is the set of $n$-grams of a text, $\ell$ is the link content, and $d$ is the PR description.

\subsection{Evaluation Dataset}
Since no existing dataset of GitHub PRs with link ranking exists, we constructed a dataset of PRs with manually annotated link relevance.
First, we filtered the GitHub dataset that we used in RQ1 to include only PRs with links in their descriptions that fall within the upper quartile of link frequency (Q3--Q4), corresponding to PRs containing 3 or more links. This threshold was determined based on the distribution observed in the methodology for RQ1: the minimum, first quartile, and second quartile were all 1 link, the third quartile was 3 links, and the maximum was 1,571 links. These statistics indicate that while most PRs contain only a small number of links, a smaller subset of PRs includes substantially higher link density and therefore presents greater contextual complexity for reviewers. By focusing on PRs in the Q3--Q4 range, the dataset emphasizes cases where link ranking is most necessary and practically beneficial.
There are 136,553 PRs, with 1,910,134 links, that fall into this range.

Following this filtering process, we selected 700 PRs (see Table~\ref{tab:dataset}) using stratified sampling across repositories to preserve the original repository-level distribution, which exceeds the statistical representative sampling (i.e., 384 with 95\% confidence level and 5\% margin of error). The resulting sample was then used for manual ranking and for constructing the link-ranking evaluation dataset.

\begin{table}[tb]
\centering
\caption{The dataset of ranked links in GitHub PRs}
\label{tab:dataset}
\begin{tabular}{rrrrr}
\toprule
No. of PRs & No. of Projects &  Min/Max/Avg. No. of Links \\
\midrule
700 & 50 & 3.0/578/14.0 \\
\bottomrule
\end{tabular}
\end{table}

Next, we performed the manual link ranking process. Two annotators (the first and third authors), with 3 years of programming experience and 1 year of GitHub pull request code review experience, worked independently. They followed the same process: reading the PR descriptions, following each link and its content, and ranking the links in each PR based on their importance for understanding the pull request's context and purpose. This instruction deliberately treats relevance as a single, role-agnostic notion of importance, i.e., how much a link helps a reviewer understand why the change was made and what it does, rather than tailoring it to a specific reviewer role or review task. We adopted this generic definition because it provides a consistent, reproducible target for training and evaluation, and because building an overall understanding of the change is the most common goal shared among reviewers. We acknowledge that relevance can also be role- and intent-dependent, and we discuss the implications of this design choice in Sections~\ref {sec:discussion} (Discussion) and~\ref {sec:threats} (Threats to Validity).

To evaluate annotators' consistency, Kendall's Tau~\cite{kendall,KendallGibbons1990} is used as the inter-rater agreement metric. The overall agreement score across 700 pull requests is 0.65, indicating moderate agreement among annotators~\cite{KendallGibbons1990}. 
When disagreements arose, annotators discussed and resolved them to produce a finalized dataset. This finalized dataset, referred to as the link-ranking dataset, is subsequently used to train the LTR model and to evaluate and compare the performance of the 5 proposed ranking approaches.

The data of 700 pull requests is split as follows: 630 PRs (90\%) are used for training the LTR model, and the remaining 70 PRs (10\%) are held out as a test set, of which 68 PRs (356 links) are used in the evaluation after removing two rollup PRs, which bundle several pull requests. The split is a simple random split at the PR level, without stratification by repository or link density. The test PRs cover 18 repositories, all of which also appear in the training set, and their numbers of annotated links per PR match those of the training set (median 4; mean 5.3 vs.\ 5.4).
To ensure a fair comparison, this same held-out set of 68 PRs is used to evaluate every ranking approach: the off-the-shelf models (TF-IDF, SBERT, the LLM-based encoders, and the cross-encoder reranker) are applied directly to it, and the trained LTR model is evaluated on it after being fit on the 630 training PRs. None of the test PRs is seen during LTR training or hyperparameter tuning, so the metrics reported in Section~\ref{sec:rq2_results} are directly comparable across models. Each approach ranks the links of a PR by its predicted relevance score; when an approach assigns the same score to several links, the tied links keep the order in which they appear in the PR description. The training set is further partitioned into 10 folds for cross-validation during hyperparameter tuning of the LTR model, as described after the next section (in Section~\ref{sec:tenfold}).

\subsection{Error Measures}

To assess ranking accuracy against the manually labeled dataset, we use four standard ranking-based information retrieval metrics. 
All metrics are computed per PR and averaged over the test PRs.
\textit{Normalized Discounted Cumulative Gain (NDCG)} sums the position-discounted relevance of the ranked links and normalizes it by the ideal ordering, yielding a value in $[0,1]$. We report NDCG@10~\cite{NDCG}, using graded relevance $n - r + 1$ for a link with ground-truth rank $r$ in a PR with $n$ links.
\textit{Precision at 1 (P@1)} is a binary measure of whether the top-ranked link is the most relevant link in the ground truth~\cite{prec}.
\textit{Mean Reciprocal Rank (MRR)} averages, across PRs, the reciprocal of the predicted position of the most relevant link, rewarding rankings that surface it early~\cite{voorhees1999mrr}.
\textit{Kendall's Tau} captures the ordinal association between the predicted and ground-truth orderings, ranging from $-1$ (perfect disagreement) to $+1$ (perfect agreement) with Tau-b to handle tied ranks~\cite{kendall,KendallGibbons1990}.

Given the size of the test set, we report 95\% confidence intervals (CIs) obtained with a percentile bootstrap over PRs (10,000 resamples)~\cite{efron1993bootstrap} and test whether the best model differs from each other approach using paired tests on the per-PR scores: a two-sided randomization test (100,000 sign flips)~\cite{smucker2007comparison} for NDCG@10, MRR, and Kendall's Tau, and an exact McNemar test~\cite{mcnemar1947note} for the binary P@1. The $p$-values are adjusted with the Holm correction~\cite{holm1979simple} across the comparisons within each metric, and differences are considered significant at $\alpha = 0.05$.

\subsection{10-Fold Cross Validation for LTR model}
\label{sec:tenfold}
While the 4 link ranking models can be used off-the-shelf, the LTR model requires training. Thus, we employed a 10-fold cross-validation strategy with group-wise splitting~\cite{crossval} to train the model and derive its hyperparameters. 
We trained a LambdaMART LTR model based on LightGBM, which used Sentence-Transformers' all-MiniLM-L6-v2 for cross-encoding and similarity extraction.
The 630 PRs were divided into ten disjoint folds. For each iteration, nine folds were used for training, and the remaining fold for validation. This process was repeated ten times, such that each fold served as the validation set exactly once. %

We chose the best hyperparameter setting based on a unified evaluation criterion, a weighted ranking score that combines NDCG@10 and MRR, focusing on the overall ranking and importance of the first-ranked result, as shown in the equation below:

\begin{equation}
\text{Score} = (0.5 \times \text{NDCG@10}) + (0.5 \times \text{MRR})
\end{equation}

We evaluate each candidate parameter configuration across the same 10-fold partitioning scheme and select the hyperparameter setting from the best-performing fold.

After completing 10-fold cross-validation, the configuration from fold 1 achieved the highest combined NDCG and MRR score (0.8292) with a relatively low standard deviation, indicating strong and stable ranking performance across query partitions. It employs the \texttt{LambdaRank} objective with GBDT boosting~\cite{gdbt}, 1,000 estimators, a learning rate of 0.005, 31 leaves, a maximum tree depth of 5, subsampling and column-sampling ratios of 0.8, regularization (\texttt{reg\_alpha}=0.05, \texttt{reg\_lambda}=2.0), and \texttt{lambdarank\_norm} enabled to normalize gradients across varying candidate-list sizes~\cite{lambdarank}. The full settings are available in our replication package~\cite{replication}.

\begin{figure}[tb]
    \centering
    \includegraphics[width=0.9\linewidth]{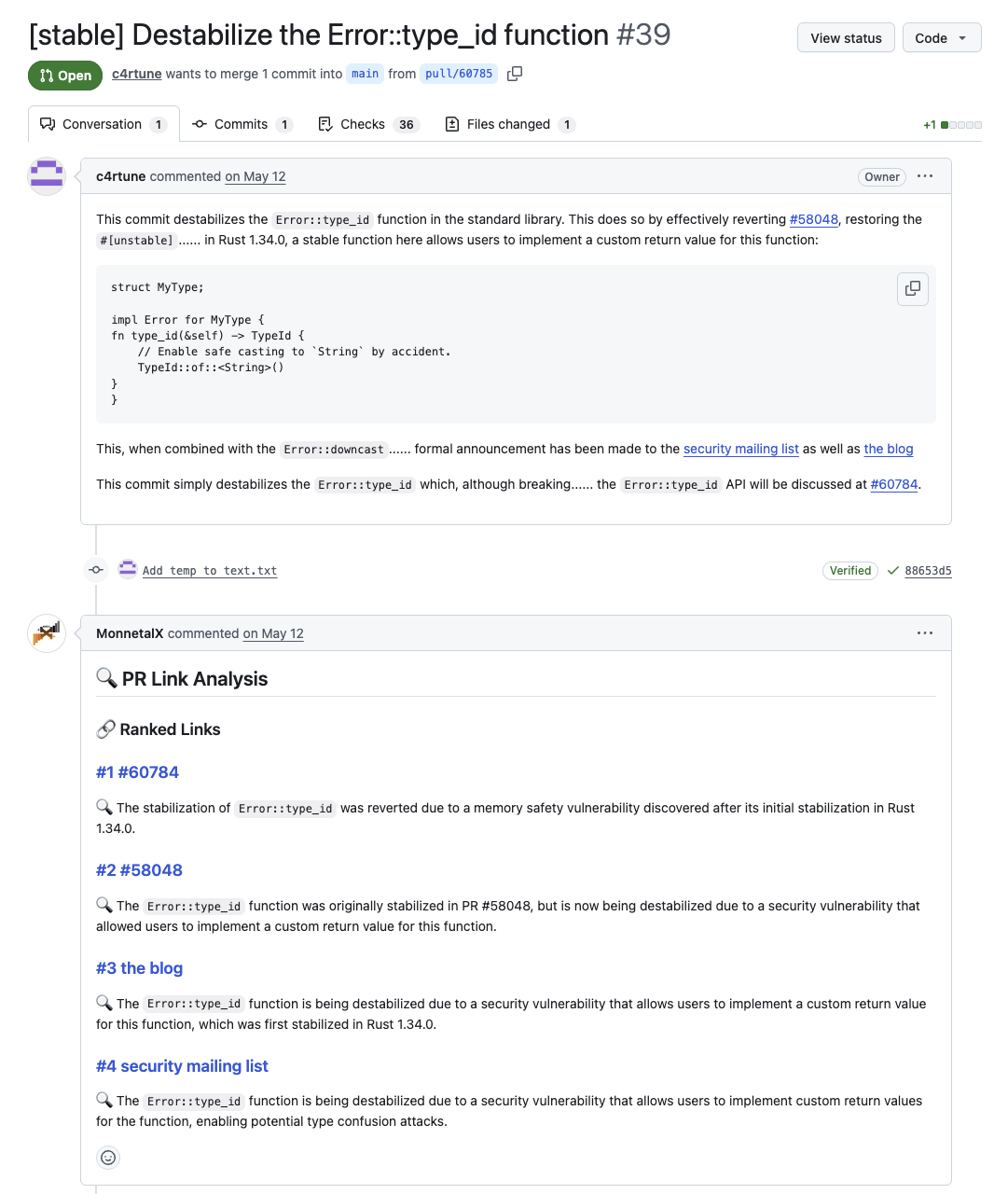}
    \caption{Link ranking and summarization implemented as a GitHub Action.}
    \label{fig:monnetalx_screenshot}
\end{figure}

\subsection{RQ3: Human Evaluation of PR Link Ranking Approach}
\label{sec:rq3_methodology}
For RQ3, we employed a human evaluation to assess the effectiveness of our proposed link-ranking approach. 
First, we implement the proposed approach as a GitHub Action tool called MonnetalX. By implementing our link ranking technique as a GitHub Action, we can seamlessly integrate it into the existing pull request review process, providing real-time feedback to reviewers as they evaluate code changes. %
We also adopt the link summarization technique proposed by Trakoolgerntong et al.~\cite{trakoolgerntong} to generate summaries for the top-ranked links, further enhancing the contextual information available to reviewers. An example of the link ranking and summarization result produced by the MonnetalX tool is illustrated in Figure~\ref{fig:monnetalx_screenshot}.

The participants were recruited via convenience sampling across multiple channels: social networks, chat applications, and email. The inclusion criteria for the participants include individuals with at least 1 year of programming experience, fluency in English, and at least 1 year of experience using GitHub, particularly GitHub pull requests.\footnote{This user study was approved by the Mahidol University Central Institutional Review Board, COA No. MU-CIRB 2026/134.3003.} These criteria ensure that participants possess sufficient background knowledge to understand the evaluation tasks and provide meaningful feedback on the proposed technique and tool.
The study was conducted online via Microsoft Teams.
Before beginning the human evaluation process, each participant received an information sheet outlining the project objectives, study procedures, and ethical considerations. Informed consent was obtained from all participants before participation in the evaluation.

\paragraph*{Evaluation Procedure}

During each session, we asked participants to review two pull requests that were selected based on two criteria: their ease of understanding of the PR description, and the number of links (4 and 8 respectively), which deemed manageable given the session's limited time.\footnote {\url{https://github.com/rust-lang/rust/pull/60785} and \url{https://github.com/electron/electron/pull/25522}}
Specifically, participants were asked to assume the role of a reviewer and thoroughly read all content, including the description and embedded links. Next, we asked them to manually rank the links in the PR description. We then introduced our tool and asked participants to compare its generated link rankings and summaries against their manual rankings and to provide feedback on the ranking results. 
Then, the participants rated the link ranking based on the following three user-centric metrics, i.e., time saving, greater focus, and reduced cognitive load, using a 5-point Likert scale (1--5). Finally, we collected open-ended feedback from participants regarding their experience with our approach and tool, as well as their suggestions for improvement.

\section{Results}
\label{sec:results}

\subsection{RQ1: How prevalent are hyperlinks in PR descriptions, and what is their distribution across different types of resources?}
\label{sec:rq1_results}

\begin{table}[tb]
\centering
\caption{Statistics of hyperlinks in PR descriptions}
\label{tab:hyperlink_stats}
\begin{tabular}{lr}
\toprule
Statistic & Values \\
\midrule
No. of PRs & 486,423 \\
Mean no. of links (SD) & 4.69 (20.50) \\
No. of links in Q1/Q2/Q3/Q4 & 1.0/2.0/3.0/1,571 \\
\bottomrule
\end{tabular}
\end{table}

After analyzing the pull requests and their links in the 50 selected GitHub projects, we found that the dataset contains 486,423 pull requests with hyperlinks, averaging 4.69 links per pull request (see Table~\ref{tab:hyperlink_stats}). However, the large standard deviation (20.50) and variance (420.36) indicate a highly skewed distribution, where a relatively small number of pull requests contain exceptionally large numbers of links. Quartile analysis further reveals that the first quartile (Q1) contains 1 link per pull request, while the second quartile (Q2) increases to 2 links, and the third quartile (Q3) reaches 4 links. The maximum observed value (in Q4) is 1{,}571 links within a single pull request. These findings suggest that although most pull requests contain only a few linked references, a subset contains substantially more links.

From Table~\ref{tab:link_types}, approximately 95.64\% of all extracted links were associated with text-based media types, followed by 2.12\% linking to Unknown (i.e., links for which the Content-type could not be determined), 1.22\% to image resources, and 0.89\% to application categories, with the remaining percentage distributed among invalid links, binary files, and videos.

To further understand the sources of the referenced content, we examined whether the links pointed to internal (i.e., GitHub) or external sources and found that 78.03\% of all extracted links were internal, referring to pull requests, issues, discussions, commits, files, or repository-level resources, while the remaining 21.97\% were external links leading to websites outside the GitHub domain. 
The most frequently referenced external domain is \texttt{docs.renovatebot.com}, which is a documentation site for Renovate Bot, a tool that automates dependency updates. Other frequently referenced domains include \texttt{our.internmc.facebook.com} and \texttt{skia.googlesource.com}, which are internal documentation and source code repositories for Facebook and Google, respectively. These findings suggest that developers often rely on both platform-native resources and external documentation or reference materials to provide context and support for their pull requests.

\begin{table}[tb]
\centering
\caption{Type of links in PR descriptions from 50 GitHub repositories.
Percentages are rounded to two decimals and may not sum to 100\%.}
\label{tab:link_types}
\begin{tabular}{lrrrr}
\toprule
Type of link & Total & Percent & Internal & External \\
\midrule
Text         & 2,182,354 & 95.64\%    & 1,754,947 & 427,407 \\
Unknown      &    48,437 &  2.12\%    &     8,454 &  39,983 \\
Images       &    27,819 &  1.22\%    &       103 &  27,716 \\
Applications &    20,411 &  0.89\%    &    16,963 &   3,448 \\
Invalid      &     2,585 &  0.11\%    &         1 &   2,584 \\
Binary       &        89 & $<$0.01\%  &         0 &      89 \\
Videos       &        36 & $<$0.01\%  &         0 &      36 \\
Others       &        25 & $<$0.01\%  &         0 &      25 \\
\midrule
Total        & 2,281,756 & 100.00\%   & 1,780,468 & 501,288 \\
\bottomrule
\end{tabular}
\end{table}

\begin{figure}[tb]
    \centering
    \includegraphics[width=0.9\columnwidth]{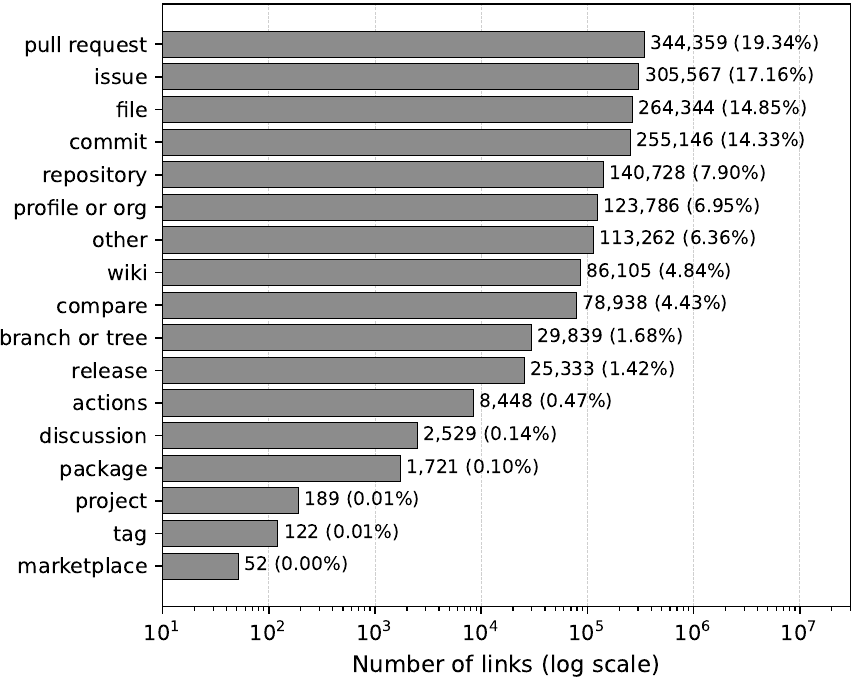}
    \caption{Types of GitHub links in extracted PRs}
    \label{fig:github_links}
\end{figure}

We also classified the GitHub-internal links by their destination using a rule-based URL classifier (Figure~\ref{fig:github_links}). Among the 1,780,468 internal links, the most frequently referenced destinations are pull requests (19.34\%), issues (17.16\%), files (14.85\%), and commits (14.33\%), reflecting the practice of cross-referencing source artifacts, code changes, and related reviews. The remaining links form a long tail across repositories, profiles, wikis, comparisons, releases, actions, discussions, branches, and packages.

To answer RQ1, we found that \textbf{hyperlinks are prevalent in GitHub PRs (4.69 links per PR on average), although the distribution is highly skewed, with some PRs containing over 1,500 links. Most links point to text-based resources (95.64\%) and are predominantly GitHub-internal references (78.03\%), dominated by pull requests, issues, files, and commits, with the remaining 21.97\% pointing to external resources.}

These findings highlight the need for automated link summarization and ranking techniques to help reviewers efficiently process pull requests with several linked references.

\subsection{RQ2: What is the effectiveness of different techniques in ranking links in PR descriptions based on their relevance?}
\label{sec:rq2_results}

\begin{table*}[t]
\centering
\caption{Comparison of Link Ranking Models: Mean [95\% Bootstrap CI] (P@1=Precision@1 and KT=Kendall's Tau).}
\label{tab:autoeval}
\small
\begin{tabular}{lcccc}
\toprule
\textbf{Model} & \textbf{NDCG@10} & \textbf{P@1} & \textbf{MRR} & \textbf{KT} \\
\midrule
\textbf{Original} & 0.915$^\dagger$ {\footnotesize[0.89, 0.93]} & 0.426$^\dagger$ {\footnotesize[0.31, 0.54]} & 0.656$^\dagger$ {\footnotesize[0.58, 0.73]} & 0.464$^\dagger$ {\footnotesize[0.33, 0.59]} \\
\midrule
\textbf{TF-IDF} & 0.917$^\dagger$ {\footnotesize[0.90, 0.94]} & 0.529 {\footnotesize[0.41, 0.65]} & 0.701 {\footnotesize[0.62, 0.78]} & 0.354$^\dagger$ {\footnotesize[0.23, 0.48]} \\
\midrule
\multicolumn{5}{l}{\textbf{SBERT Encoder}} \\
all\_minilm\_l6\_v2 & 0.938 {\footnotesize[0.92, 0.95]} & 0.588 {\footnotesize[0.47, 0.71]} & \textbf{0.756} {\footnotesize[0.68, 0.83]} & 0.429$^\dagger$ {\footnotesize[0.32, 0.53]} \\
\midrule
\multicolumn{5}{l}{\textbf{Reranker (Cross-Encoder)}} \\
jina\_reranker & 0.924 {\footnotesize[0.91, 0.94]} & 0.485 {\footnotesize[0.37, 0.60]} & 0.704 {\footnotesize[0.63, 0.78]} & 0.400$^\dagger$ {\footnotesize[0.28, 0.51]} \\
\midrule
\multicolumn{5}{l}{\textbf{LLM as Encoders}} \\
deepseek\_coder & 0.924 {\footnotesize[0.90, 0.94]} & 0.529 {\footnotesize[0.41, 0.65]} & 0.693 {\footnotesize[0.61, 0.77]} & 0.361$^\dagger$ {\footnotesize[0.24, 0.48]} \\
mistral7B & 0.932 {\footnotesize[0.92, 0.95]} & 0.544 {\footnotesize[0.43, 0.66]} & 0.720 {\footnotesize[0.65, 0.79]} & 0.429$^\dagger$ {\footnotesize[0.32, 0.53]} \\
\midrule
\multicolumn{5}{l}{\textbf{Learning-to-rank (LTR)}} \\
LightGBM-4-features & 0.901$^\dagger$ {\footnotesize[0.88, 0.92]} & 0.353$^\dagger$ {\footnotesize[0.24, 0.47]} & 0.606$^\dagger$ {\footnotesize[0.53, 0.68]} & 0.331$^\dagger$ {\footnotesize[0.18, 0.48]} \\
LightGBM-8-features & \textbf{0.952} {\footnotesize[0.94, 0.97]} & \textbf{0.618} {\footnotesize[0.50, 0.74]} & 0.752 {\footnotesize[0.67, 0.83]} & \textbf{0.627} {\footnotesize[0.53, 0.72]} \\
\bottomrule
\end{tabular}
\end{table*}

The results shown in Table~\ref{tab:autoeval}, computed on the same 68 test PRs for each model, indicate that the LTR model achieves the best overall performance among the evaluated approaches. In particular, the LightGBM model trained with eight features (defined in Section~\ref{sec:ltr}) obtained the highest NDCG@10 score of 0.952, the highest Precision@1 score of 0.618, and the highest Kendall's Tau score of 0.627, while also achieving an MRR of 0.752, close to the best MRR of 0.756 obtained by SBERT.
The paired significance tests show where this advantage is reliable; in Table~\ref{tab:autoeval}, $^\dagger$ marks values that are significantly lower than those of LightGBM-8-features (Holm-corrected $p < 0.05$).The model's rankings match the manually assigned ground-truth order significantly better than those of every other approach: its Kendall's Tau is significantly higher than that of TF-IDF, Sentence-BERT, the cross-encoder reranker, the LLM-based encoders, the 4-feature LTR model, and the Original ordering (Holm-corrected $p \le 0.02$). On the metrics that focus on the top of the ranking, NDCG@10, Precision@1, and MRR, it significantly outperforms the Original ordering and the 4-feature LTR model, and TF-IDF on NDCG@10, whereas its differences from Sentence-BERT, the cross-encoder reranker, and the LLM-based encoders are not statistically significant on the 68 test PRs. The 4-feature LTR model is the weakest approach on all four metrics, showing that link content length and $n$-gram lexical overlap are essential to the LTR model's performance.

Looking at the Original baseline, which preserves the order in which the PR author placed the links in the description, we found that its Kendall's Tau (0.464) exceeds that of every non-LTR automated approach (0.354--0.429), and its NDCG@10 (0.915) is close to theirs. This indicates that authors do not order links arbitrarily. The as-written order already carries a meaningful signal about the overall relevance order, making it a non-trivial baseline to beat. However, the author's first link is often not the most relevant one: the Original ordering has the lowest Precision@1 (0.426) and MRR (0.656) among the non-LTR approaches. The LightGBM model with 8 features significantly improves over the Original ordering on all four metrics, with the largest gains on Precision@1 (by 44.8\%; it ranks the most relevant link first in 17 PRs where the Original ordering does not, against 4 PRs where only the Original ordering does) and Kendall's Tau (by 35.2\%), alongside gains of 14.8\% in MRR and 4.1\% in NDCG@10. In practice, although authors impose a partial structure, many highly relevant links remain buried later in the description (as in the motivating example of Figure~\ref{fig:example}). LTR reorders them toward the top, giving reviewers a measurable improvement over the default.

To understand which signals drive the model, we applied SHAP~\cite{lundberg2017unified} to its predictions on the held-out set. Cosine similarity between the PR description and the candidate link is by far the most influential feature (mean $|\text{SHAP}|=0.47$), contributing roughly as much as all others combined, followed by link content length ($0.17$) and unigram overlap ($0.10$). The model's ranking quality is thus primarily attributable to content's semantic similarity, complemented by link content length and lexical overlap.

To answer RQ2, we found that \textbf{
LTR with 8 lexical, semantic, and structural features (LightGBM) achieves the highest NDCG@10, Precision@1, and Kendall's Tau. It significantly outperforms the PR author's original link ordering on all four metrics and orders links significantly closer to the ground truth than all other approaches, while performing on par with semantic encoders and the cross-encoder reranker in placing the most relevant link at the top.}

\subsection{RQ3: How do software developers perceive the effectiveness of the link ranking approach during PR reviews?}
\label{sec:rq3_results}

\begin{figure}[tb]
    \centering
    \includegraphics[width=\columnwidth]{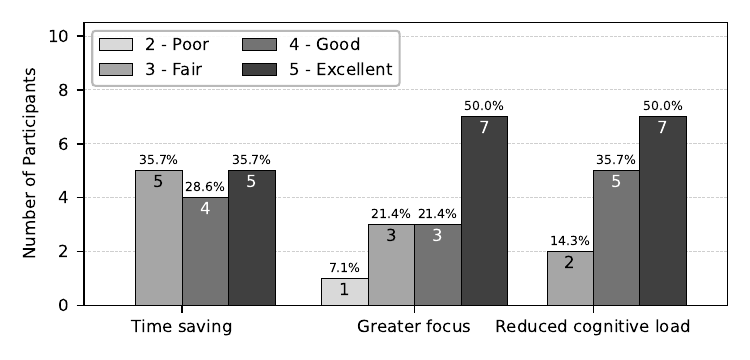}
    \caption{Human evaluation results for the link ranking approach.}
    \label{fig:ranking_score}
\end{figure}

For this RQ, we recruited 10 software practitioners and 4 graduate students.
The human evaluation of the link prioritization mechanism was assessed using three user-centric metrics: time saving, greater focus, and reduced cognitive load. 
The quantitative feedback from the 14 participants, as illustrated in \autoref{fig:ranking_score}, demonstrates the approach's real-world effectiveness in enhancing the pull request review process.

Using a 5-point Likert scale, participants rated all three aspects positively. \textit{Reduced cognitive load} scored highest, with 85.7\% (12 of 14) rating it Good (7 participants) or Excellent (5) and none below Fair (2). \textit{Greater focus} followed, with 71.4\% (10 of 14) rating it Good or Excellent, seven of them Excellent, though one participant rated it Poor (2). \textit{Time saving} received the most modest ratings, with 64.3\% (9 of 14) Good or Excellent, and the remaining 35.7\% (5 of 14) Fair. These results indicate that presenting prioritized links helps surface relevant information more quickly, focus on key resources, and reduce the effort required to navigate dispersed context.

We also collected qualitative feedback. Since linking to issue reports in PR descriptions is a best practice~\cite{Jiang2026}, participants found the ranking useful for cutting unnecessary information, but suggested it better align with developer investigation workflows. Several preferred a ``problem-first'' ordering that surfaces the links to the root-cause issue before supporting documentation. Participants also noted that what counts as important is role-dependent, e.g., some prioritized technical documentation while others preferred issue or bug reports first, and suggested separating low-priority resources, such as contribution guidelines, into a distinct section to reduce distraction.

To answer RQ3, we found that \textbf{software developers perceive the link ranking approach as beneficial, most strongly in reducing cognitive load (85.7\% Good or Excellent), followed by improving focus (71.4\% Good or Excellent), and saving review time (64.3\% Good or Excellent). Qualitatively, participants valued the prioritization but indicated that relevance is role- and goal-dependent.}

\section{Discussion}
\label{sec:discussion}

In this section, we interpret our findings across the three research questions, relate them to prior work, and discuss their implications for researchers and practitioners.

\subsection{Links are a pervasive but skewed source of review context}
Our preliminary study (RQ1) confirms that hyperlinks are a routine part of how developers communicate context in pull requests, averaging approximately 4.9 links per PR across nearly 500{,}000 PRs. This complements Wang et al.~\cite{Wang2021}, who reported that roughly one-fourth of code review discussions contain at least one shared link, and extends their OpenStack/Qt observation to the GitHub PR ecosystem at a substantially larger scale. More importantly, the distribution is highly skewed: while the second-quartile PR contains only two links, the upper quartile and long tail (up to 1{,}571 links in a single PR) indicate that a non-trivial subset of PRs imposes a disproportionately high contextual burden on reviewers, precisely where a link ranking and summarizing tool yields the greatest practical benefit.
The dominance of GitHub-internal links (78.03\%) over external links (21.97\%), together with the predominance of commit, PR, file, and issue references among internal links, further suggests that an effective tool must handle heterogeneous link types rather than specializing in a single resource category.

\subsection{Link ranking needs lexical and semantic features}
The ranking evaluation (RQ2) shows that our LightGBM LambdaMART model with eight features achieves the strongest overall performance: its link ordering matches the ground truth significantly better than those of TF-IDF, SBERT, cross-encoder rerankers, LLM-based encoders, and the PR author's own link ordering, and it significantly outperforms the author's ordering on all four metrics. The significant gain from the four- to the eight-feature configuration on all four metrics, driven by link content length and $n$-gram lexical overlap, shows that lexical evidence remains valuable even alongside strong semantic embeddings: the most effective approach is a learned combination of lexical, semantic, and structural signals rather than a purely semantic one. This is why a lightweight gradient-boosted model matches or surpasses larger neural encoders, making a supervised model trained on a modestly annotated dataset both accurate and economical for relevance ranking in this setting.

\subsection{Perceived benefits with workflow mismatches}
The human evaluation (RQ3) results are encouraging but uneven across the three aspects. Cognitive load shows the strongest signal (85.7\% Good or Excellent, none below Fair), suggesting that presenting prioritized links directly within the PR effectively eases navigation of dispersed context. Improved focus follows closely, although one participant rated it Poor, indicating that a ranking that does not match a reviewer's expectation can distract rather than help. Time saving received the lowest ratings. A plausible explanation is that reviewers still open and verify the top-ranked links, so inspection time shifts rather than disappears. The benefit is felt more in reduced mental effort than in raw speed.

At the same time, the qualitative feedback reveals a gap between optimizing links' relevancy and supporting how developers actually investigate changes. Participants expressed a preference for a ``problem-first'' ordering, in which links to root-cause issues or bug reports are surfaced before supporting documentation, and several noted that the definition of relevance is role-dependent: a reviewer focused on correctness may prioritize issue reports, whereas one focused on API design may prioritize technical documentation. This suggests that contextual relevance, as captured by our current features, is necessary but not sufficient, and that per-project or per-role customization may be needed. Link ranking would benefit from incorporating link intent (e.g., distinguishing problem statements from reference material, as in the link-intention taxonomy of Wang et al.~\cite{Wang2021}) and from letting reviewers express role-specific preferences. The request to separate low-priority resources, such as contribution guidelines, into a distinct section reinforces that presentation, not only ordering, also matters.

\subsection{Implications}
For researchers, our results indicate that link ranking in code review is best framed as a learning-to-rank problem fusing lexical, semantic, and structural features rather than a pure embedding-similarity task, with intent- and role-aware relevance a promising direction for future modeling. For practitioners, MonnetalX shows that meaningful reductions in review effort can be delivered through a lightweight model deployed natively as a GitHub Action, without additional infrastructure. The gap between offline measures and human evaluation also argues for keeping the human in the loop, e.g., by exposing configurable ranking objectives and grouping links by intent.

\section{Threats to Validity}
\label{sec:threats}

\textbf{Internal validity:} The dataset was annotated by the research team, introducing potential subjectivity. We mitigated this with double-annotation, a conflict-resolution phase, and inter-rater agreement. The human evaluation included 14 participants, potentially limiting statistical power. Likewise, the RQ2 test set contains 68 PRs; we therefore report bootstrap confidence intervals and Holm-corrected paired significance tests, under which some differences among the best-performing approaches are not statistically significant.
To accommodate participants' time constraints, our evaluation used PRs with lower link densities (4 and 8 links), which may not fully capture developer behavior under extreme information overload (e.g., $>$50 links).

\textbf{External validity:} The preliminary study covers 50 repositories selected by star rating, which may skew toward mature, well-maintained projects. The 700-PR dataset focuses on high-link-density PRs. Thus, the results may not fully generalize to low-link PRs. Since the train/test split is random at the PR level, all test PRs come from repositories that also appear in the training data, so the LTR model's generalization to unseen repositories is not evaluated.The tool was evaluated only on public repositories. Performance on private or enterprise repositories with different link patterns may differ.

\textbf{Construct validity:} NDCG, MRR, P@1, and Kendall's Tau are standard information retrieval metrics but may not capture all reviewer preferences. The 5-point Likert scales are intuitive but not validated instruments, which may limit comparability. Our ground truth relies on a single global relevance ordering per PR, which may not fully capture role-specific relevance. %

\section*{Acknowledgments}
We gratefully acknowledge the financial support of: 1) JSPS for the KAKENHI grants (JP25K22845, JP26H02500, JP26K21198); (2) Japan Science and Technology Agency (JST) as part of Adopting Sustainable Partnerships for Innovative Research Ecosystem (ASPIRE), Grant Number JPMJAP2415, (3) the Kayamori Foundation of Informational Science Advancement for supporting
Tao Xiao, and (4) the Inamori Research Institute for Science for supporting Yasutaka Kamei via the InaRIS Fellowship.

\section{Conclusion}
\label{sec:conclusion}

This paper presents an automated approach that ranks hyperlinks in PR descriptions to reduce reviewer cognitive load. A preliminary study of 486,423 PRs across 50 repositories confirmed that hyperlinks are pervasive in code review and that their distribution is highly skewed, motivating automated support for PRs with dense contextual references. By framing link ranking as a learning-to-rank problem, our LightGBM model with eight lexical, semantic, and structural features achieves the best ranking performance, ordering links significantly closer to the ground truth than the original ranking, TF-IDF, LLM-based encoders, and cross-encoder rerankers. A human evaluation with 14 participants confirms the practical benefits of the approach: participants reported that the ranked links help them save review time, focus on important resources, and reduce cognitive load, with 85.7\% rating the reduction in cognitive load as ``Good'' or ``Excellent.'' In future work, we plan to incorporate link intent and role-specific reviewer preferences into the ranking and to group links by purpose, so that the ordering better matches developer investigation workflows. Our prototype tool, implementing the approach as a GitHub Action, MonnetalX, is publicly available on GitHub and the GitHub Marketplace~\cite{monnetalx}. The replication package for this study is publicly available~\cite{replication}.

\bibliographystyle{IEEEtran}
\bibliography{references}

\end{document}